%% file: paper.tex
\documentclass[conference]{IEEEtran}

\usepackage{graphicx}
\usepackage{subfig}
\usepackage{listings}
\usepackage{enumitem}
\usepackage{todonotes}

\graphicspath{{../images/}}
\DeclareGraphicsExtensions{.pdf,.jpeg,.png,.jpg,.eps}

\begin{document}
\title{The BigDAWG Polystore System and Architecture}



\author{\IEEEauthorblockN{Vijay
    Gadepally\IEEEauthorrefmark{1}\IEEEauthorrefmark{2} Peinan
    Chen\IEEEauthorrefmark{2}  Jennie
    Duggan\IEEEauthorrefmark{3}  Aaron Elmore\IEEEauthorrefmark{4}
    Brandon Haynes\IEEEauthorrefmark{6}  \\ Jeremy Kepner\IEEEauthorrefmark{1}\IEEEauthorrefmark{2}
    Samuel Madden\IEEEauthorrefmark{2} Tim
    Mattson\IEEEauthorrefmark{5} Michael
    Stonebraker\IEEEauthorrefmark{2}} \\
\IEEEauthorblockA{\IEEEauthorrefmark{1}MIT Lincoln Laboratory
  \IEEEauthorrefmark{2}MIT CSAIL \IEEEauthorrefmark{3}Northwestern
  University \IEEEauthorrefmark{4}University of Chicago \\
  \IEEEauthorrefmark{5}Intel Corporation \IEEEauthorrefmark{6}University of Washington}}

\maketitle

\let\thefootnote\relax\footnotetext{The corresponding author, Vijay Gadepally, can be reached at
 \\ vijayg [at] ll.mit.edu}

\begin{abstract}

Organizations are often faced with the challenge of providing data
management solutions for large, heterogenous datasets that may have
different underlying data and programming models. For example, a medical dataset
may have unstructured text, relational data, time series waveforms and
imagery. Trying to fit such datasets in
a single data management system can have adverse performance and efficiency
effects. As a part of the Intel Science and Technology Center on Big
Data, we are developing a polystore system designed for such
problems. BigDAWG (short for the Big Data Analytics Working Group) is a polystore system designed to work on complex problems
that naturally span across different processing or storage
engines. BigDAWG provides an architecture that supports diverse
database systems working with different data models, support for the
competing notions of location transparency and semantic completeness
via \textit{islands} and a middleware that
provides a uniform multi--island interface. Initial results from a
prototype of the BigDAWG system applied to a medical dataset
validate polystore concepts. In this article, we will describe
polystore databases,
the current BigDAWG architecture and its application on the MIMIC II
medical dataset, initial performance results and our future
development plans.
\end{abstract}


\IEEEpeerreviewmaketitle

\section{Introduction}

Enterprises today encounter many types of databases, data, and
storage models. Developing analytics and applications that
work across these different modalities is often limited by the
incompatibility of systems or the difficulty of creating new
connectors and translators between each one. For example, consider the
MIMIC II
dataset~\cite{saeed2011multiparameter} which contains deidentified
health data
collected from thousands of critical care patients in an Intensive
Care Unit (ICU). This publicly available dataset (\textit{http://mimic.physionet.org/})
contains structured data such as demographics and medications; unstructured text
 such as doctor and nurse reports; and time--series data
of physiological signals such as vital signs and electrocardiogram
(ECG). Each of these components of the dataset can be efficiently organized into
database engines supporting different data models. 
For example, the structured data
in a relational database, the text notes in a key-value or graph
database and the time--series data in an array database.
Analytics of the future will cross the boundaries of a single data
modality, such as
correlating information from a doctor's note against the physiological
measurements collected from a particular sensor. Further, the same
dataset may be stored in different data engines and leveraged based on
the data engine that provides the highest performance response to a
particular query.

Such analytics on complex datasets call for the development of a 
new generation of federated databases that support seamless access to 
the different data models of database or storage engines. We 
refer to such a system as a \textit{polystore}
in order to distinguish it from traditional federated
databases that largely supported access to multiple
engines using the same data model.

As a part of the Intel Science and Technology Center (ISTC) on Big
Data, we are developing the BigDAWG, short for Big Data Analytics
Working Group, polystore system. The BigDAWG stack is designed to support 
multiple
data models, real-time streaming analytics, visualization interfaces, and multiple databases. The
current version of BigDAWG~\cite{elmore2015demonstration} shows
significant promise and has been used to develop a series of
applications for the MIMIC II dataset. The BigDAWG system supports
multiple data stores; provides an abstraction of data and programming
models through ``islands''; a middleware and API that
can be used for query planning, optimization and execution; and
support for applications, visualization and clients. Initial results
of applying the BigDAWG system to diverse data such as medical
imagery or clinical records has shown the value of a polystore system
in developing new solutions for complex data management.

The remainder of the article is organized as
follows: Section~\ref{polystore} expands on the concept of a polystore
databases and the execution of polystore queries. Section~\ref{architecture} describes the current BigDAWG
architecture and its application to the MIMIC II
dataset. Section~\ref{apps} describes performance results on an
initial BigDAWG implementation. Finally, we conclude and discuss future
work in Section~\ref{conc}.

\section{Polystore Databases}
\label{polystore}

With the increased interest in developing storage and management
solutions for disparate data sources coupled with our 
belief that 
``one size does not fit all''~\cite{stonebraker2005one},
there is a renewed interest in developing
database management systems (DBMSs) that can support multiple query
languages and complete functionality of underlying database
systems. Prior work on federated databases such as Garlic~\cite{carey1995towards}, IBM DB2~\cite{gassner1993query} and
others~\cite{sheth1990federated} have demonstrated the ability to
provide a single interface to disparate DBMSs. Other related work in parallel
databases~\cite{dewitt1992parallel} and
computing~\cite{hudak2009computational,samsi2010matlab} have demonstrated the high
performance that can be achieved by making use of replication,
partitioning and horizontally scaled hardware. Many of the federated
database technologies concentrated on relational data. With the influx
of different data sources such as text, imagery, and video, such
relational data models may not support high performance ingest and
query for these new data modalities. Further, supporting the types of analytics that users wish to
perform (for example, a combination of convolution of time series data, gaussian
filtering of imagery, topic modeling of text,etc.) is difficult within a
single programming or data model.

\begin{figure}[t!]
\centerline{
\includegraphics[width=3.6in]{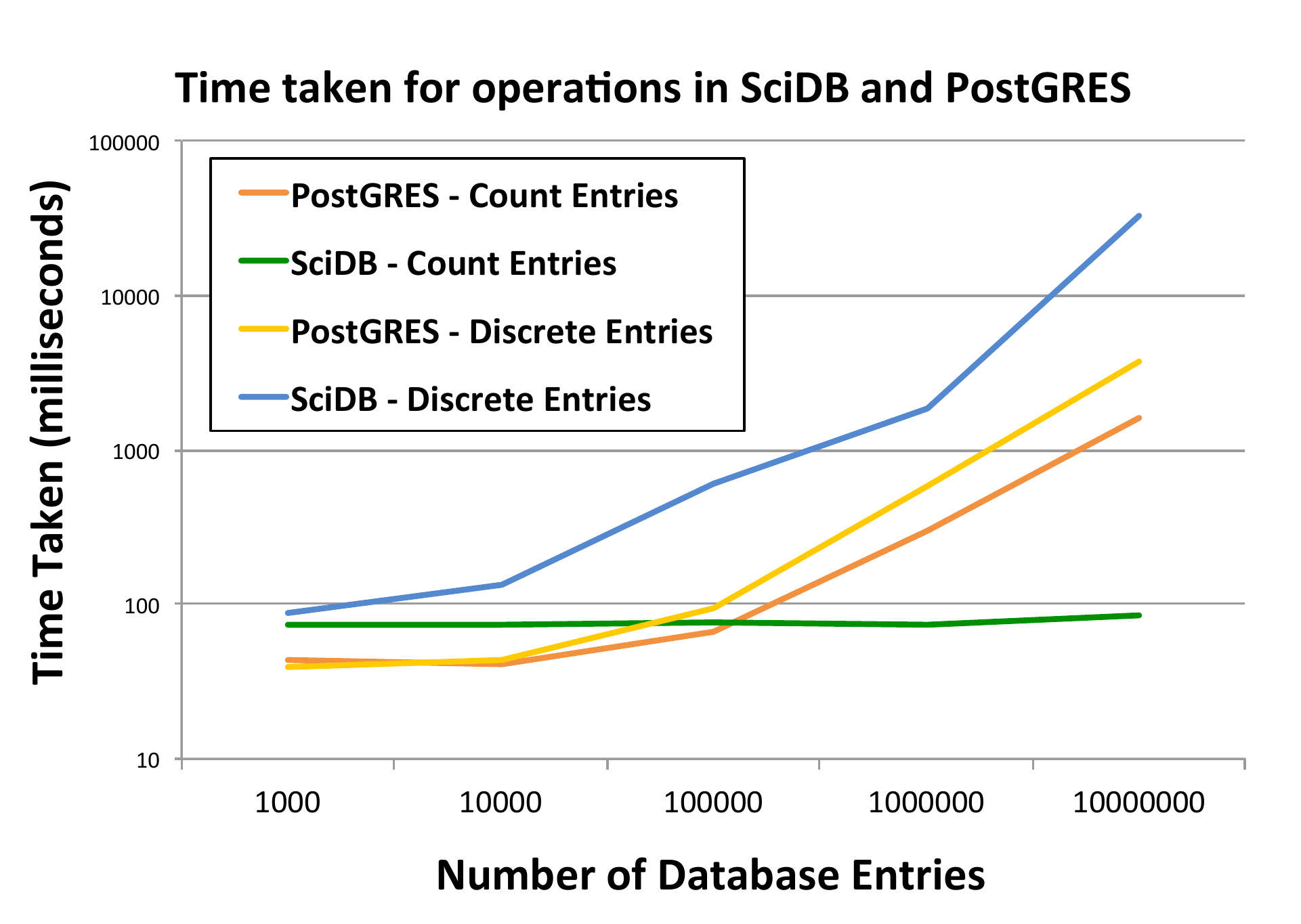}
}
\caption{Time taken for various database operations in difference
  database engines. The dashed lines correspond to a count operation
  in SciDB and PostGRES and the solid lines correspond finding the
  number of discrete entries in SciDB and PostGRES. For the count
  operation, SciDB outperforms PostGRES whereas PostGRES outperforms
  SciDB for finding the number of discrete entries.}
\label{fig:dboperations}
\end{figure}

Consider the simple
performance curve of Figure~\ref{fig:dboperations} which describes an
experiment where we performed two basic operations -- counting the number of entries
and extracting discrete entries -- on a varying number of
elements. As shown in the figure, for counting the number of entries,
SciDB outperforms PostGRES by nearly an order of magnitude. We see the
relative performance reversed in the case of extracting discrete entries.

Many time-series, image or video storage systems are most efficient
when using an array data model~\cite{cudre2009demonstration} which provides a natural
organization and representation of data. Analytics on these data are often developed
using linear algebraic operations such as matrix multiplication. In a
simple experiment in which we performed matrix multiplication in
PostGRES and SciDB, we observed nearly three orders of magnitude
difference in performance time (for a $1000 \times 1000$ dense matrix multiplication,
PostGRES takes approximately 166 minutes vs. 5 seconds in SciDB).

These results suggest that analytics in which one wishes to
perform a combination of operations (for example, extracting the
discrete entries in a dataset and using that result to perform a
matrix multiplication operation) may benefit from performing
part of the operation in PostGRES (extracting discrete entries) and the
remaining part (matrix multiplication) in SciDB.

% \begin{figure}[b!]
% \centerline{
% \includegraphics[width=3.5in]{images/matmul.pdf}
% }
% \caption{Time taken for multiplying two dense matrices of varying size
%   using PostGRES and SciDB. The x-axis indicates the total number of
%   elements and the y-axis indicates the time taken for the operation.}
% \label{fig:matmul}
% \end{figure}

Extending the concept of federated and parallel databases, we propose
a ``polystore'' database. Polystore databases can harness the relative strengths of underlying
DBMSs. Unlike federated or parallel databases, polystore databases are
designed to simultaneously work with disparate database engines
and programming/data models while supporting complete functionality of
underlying DBMSs. In fact, a polystore solution may include federated
and/or parallel databases as a part of the overall solution stack. In a
polystore solution, different components of an overall dataset can be
stored in the engine(s) that will best support high performance
ingest, query and analysis. For example, a dataset with structured,
text and time-series data may simultaneously leverage relational,
key-value and array databases. Incoming queries may leverage one or
more of the underlying systems based on the characteristics of the
query. For example, performing a linear algebraic operation on
time-series data may utilize just an array database; performing a
join between time-series data and structured data may leverage
array and relational databases respectively.

\begin{figure}[t!]
\centerline{
\includegraphics[width=3.4in]{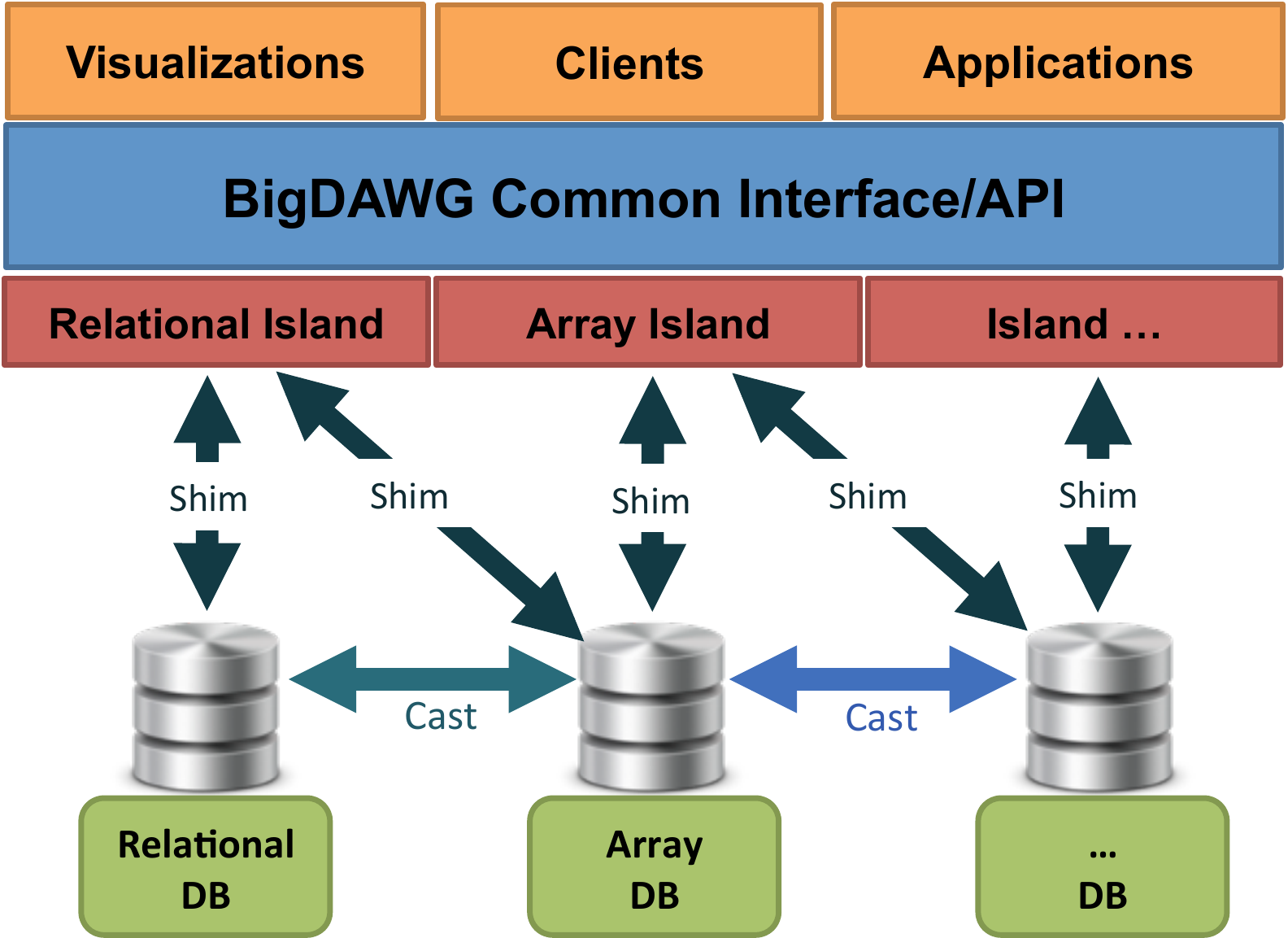}
}
\caption{The BigDAWG polystore architecture consists of four layers -
  engines, islands, middleware/API and applications.}
\label{fig:bigdawgarch}
\end{figure}

In order to support such expansive functionality, the BigDAWG
polystore system (Figure~\ref{fig:bigdawgarch}) utilizes a number of
features. ``Islands'' provide users with a number of
programming and data model choices; ``Shim'' operations allow
translation of one data model to another; and ``Cast'' operations
allow for the migration of data from one engine or island to
another. We go into greater depth of the BigDAWG architecture in Section~\ref{architecture}.

\section{BigDAWG Architecture}
\label{architecture}
The BigDAWG architecture consists of four distinct layers as described
in Figure~\ref{fig:bigdawgarch}: database and storage engines; islands; 
middleware and API; and applications. In this section, we discuss the
current status of each of these layers as well as how they are used
with the MIMIC II dataset.

\subsection{Database and Storage Engines}

A key design feature of BigDAWG is the support of multiple
database and storage engines. With the rapid increase in heterogenous
data and proliferation of highly specialized, tuned and hardware
accelerated database
engines, it is important the BigDAWG support as many 
data models as possible. Further, many organizations already 
rely on legacy systems as a part of their overall solution.  
We believe that analytics of the future
will depend on many, distinct data sources that can be
efficiently stored and processed only in disparate systems. BigDAWG is designed to
address this need by leveraging many vertically-integrated data
management systems.

For the MIMIC II dataset, we use the relational databases PostgreSQL and
Myria~\cite{halperin2014demonstration} to store clinical data such
as demographics and medications. BigDAWG uses the
key-value store Apache Accumulo for freeform text data and to perform graph analytics~\cite{hutchison2015graphulo}. For the historical
waveform time-series data of various physiological
signals, we use the array store
SciDB~\cite{stonebreaker2013}. Finally, for streaming time-series data,
our application uses the streaming database S-Store~\cite{cetintemel2014s}.

\subsection{Islands}

The next layer of the BigDAWG stack is its \textit{islands}. Islands 
allow users to trade off between semantic
completeness (using the full power of an underlying database engine)
and location transparency (the ability to access data without
knowledge of the underlying engine). Each island has a data model, a
query language or set of operators and one or more database
engines for executing them. In the BigDAWG prototype, the user determines
the {\it scope } of their query by specifying an island 
within which the query will be executed.
 Islands are a user-facing abstraction, and they are designed to reduce the challenges associated with incorporating a new database engine.

We currently support a number of islands. For example, the D4M island
provides users with an associative array data model~\cite{vijay2015} to PostgreSQL,
Accumulo, and SciDB. The Myria island exposes support for iteration over and
efficient casting between the MyriaX, PostgreSQL and SciDB databases. We also support a number of
\textit{degenerate} islands that connect to a single database
engine. These \textit{degenerate} islands provide support for
the full semantic power (programming and data model) of a connected database at the expense of
location transparency.

\subsection{BigDAWG Layer}

The BigDAWG middleware consists of a number of components required to
support the multiple islands, programming languages and
query types that BigDAWG supports.

\subsubsection{BigDAWG middleware}

\begin{figure}[t!]
\centerline{
\includegraphics[width=3.4in]{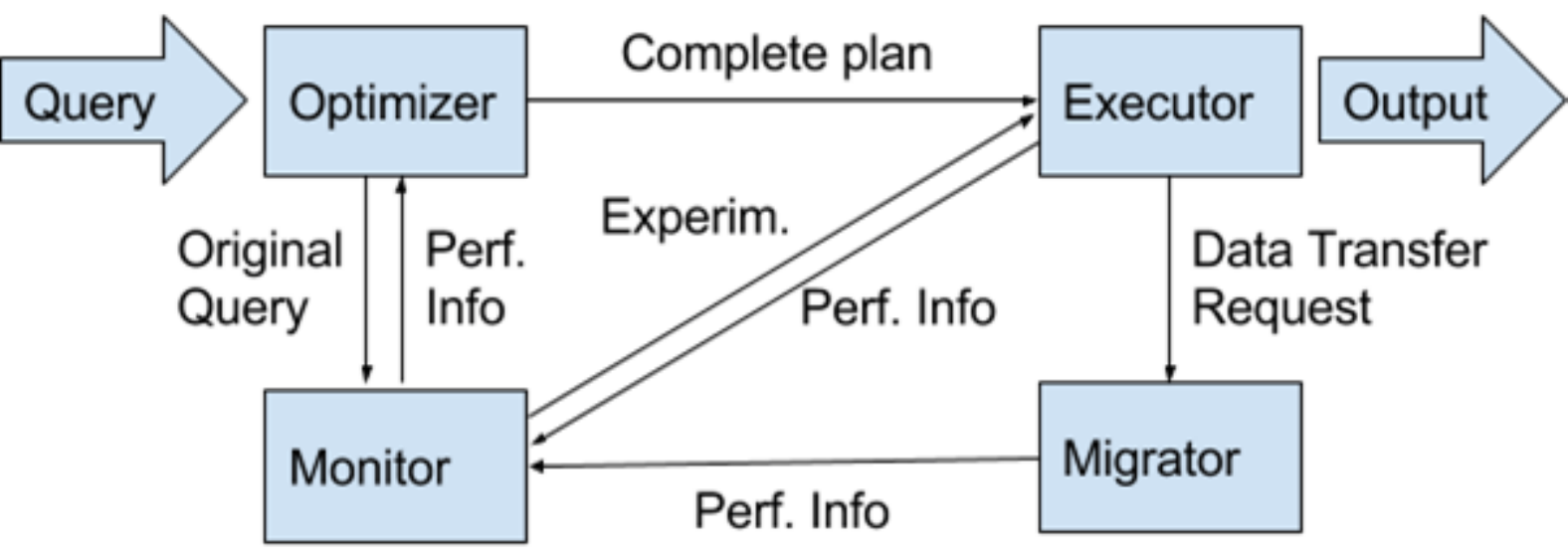}
}
\caption{The BigDAWG middleware consists of four modules - planner,
  monitor, migrator and executor. For a new query, it is first passed
  to the planner which interacts with the monitor to develop a
  complete query plan. This is then passed to the executor which
  leverages the migrator as needed to complete the query.}
\label{fig:bigdawgmiddlware}
\end{figure}

The BigDAWG middleware, is
responsible for receiving queries, query planning, determining
efficient execution strategies, maintaining history of previous
queries, and maintaining a record of previous query performance. The
architecture of the middleware is shown
in Figure~\ref{fig:bigdawgmiddlware}. The middleware has four
components: the query planning module (planner)~\cite{gupta2016}, the performance
monitoring module (monitor)~\cite{chen2016}, the data migration module (migrator)~\cite{dziedzic2016} and
the query execution module (executor)~\cite{she2016}. Given an incoming query, the
planner parses the query into collections of objects and creates a set of possible query plan
trees that also highlights the possible engines for each collection of
objects. The planner then sends these trees to the monitor which uses
existing performance information to determine a tree with the best
engine for each collection of objects (based on previous experience
of a similar query). The tree is then passed to the executor which
determines the best method to combine the collections of objects and
executes the query. The executor can use the migrator to move objects
between engines and islands, if required, by the query plan.

\subsubsection{BigDAWG API}

The BigDAWG interface provides a simple API to execute polystore
queries. The API layer consists of server and client facing
components. The server components incorporate the 
many possible islands which connect
to database engines via lightweight connectors referred to as
\textit{shims}. Shims essentially act as an adapter to go from the
language of an island to the native language of an underlying database
engine. In order to specify how a user is interacting with an island,
a user specifies a scope in their query. A scope of a query allow an
island to correctly interpret the syntax of the query and allows the
Island to select the correct Shim that is needed to execute a part of
the query. Thus, a cross-island query may involve multiple Scope
operations. For example, let us suppose we have two tables $A$ and $B$
in a relational and array database, respectively. Suppose that we want
to perform the cross-island query
\texttt{ARRAY(multiply(RELATIONAL(select * from A,...),B)} which
takes all the data in table $A$ and multiplies it with all the data in
table $B$. In this case, the inner operation
(\texttt{RELATIONAL(...)}) invokes the Relational scope
and the outer operation invokes the Array scope. 
Moving the data
between two engines can be done through the Cast operation. 
The Cast operation sends information about the translation between 
data models and moves the data as needed.
In the example query, this may imply that the results of
\texttt{RELATIONAL(...)} along with translational information
about the resulting objects
are Cast to the engine where
$B$ resides.

\subsubsection{Polystore Queries}
\label{querymechanics}

Efficient query execution is a key goal of the BigDAWG
system.  A key challenge is that the data being queried is likely to
be distributed among two or more disparate data management systems.
In order to support different islands, efficient data movement is also
critical. Moreover, efficient execution may also depend on system parameters
such as available resources or usage that are prone to change. To
illustrate the mechanics of a polystore query, in this section we describe the simplest case where there is no replication, partitioned objects,
expensive queries or attempts to move objects for load
balancing. Given an incoming query, an
execution plan for the query is based on whether
the query is in a training or production phase.

The training phase is typically used for execution of queries that are
new (either the query is new or the system has changed significantly
since the last time a particular query was run) or are believed to
have been poorly executed. In the simplest case, the training phase consists of queries that
arrive with a ``training'' tag. In the training phase, we allow the
query execution engine to generate a good query plan using any number
of available resources. First, the query planner parses the query and 
assigns the scope of each piece of the query to a particular
island. Pieces of the resulting subquery that are local to a particular storage
engine are encapsulated into a container and given an identifying
signature. For the remaining elements of the query (remainder), which correspond
to cross-system predicates, we generate a signature by looking at
the structure of the remainder, the objects being referenced and the
constants in the query. If the remainder signature has been seen
before, a query plan can be extracted. If not, the system decomposes
the remainder to determine all possible query plans which are then
sent to the monitor.

To execute the query, the monitor feeds the queries to the executor,
plus all of the containers which are then passed to the appropriate
underlying storage engine. For the cross-engine predicates, the
executor decides how to perform each step. The executor runs each
query, collects the total running time and other usage statistics and
stores the information in the monitor database. This information can
then be used to determine the best query plan in the production phase.

In the production phase, when a query is received it is first matched
against the various signatures in the monitor database and the optimizer selects the closest
one. The BigDAWG optimizer also compares the current usage statistics of the system
and compares it against the usage statistics of the system when the
training was performed. If there are large differences, the optimizer
may select an alternate query plan that more closely resembles the
current resources or system usage or recommend that the user rerun the
query under
the training phase under the current usage. In cases where the
signature of the incoming query do not match with existing
signatures, the optimizer may suggest the query run in training mode
or construct a list of plans as done in the training phase and have
the monitor pick one at random. The remaining plans can then be run in
the background of the system when it is underutilized. Over time,
these plans are then added to the monitor database.

\subsection{Applications and Visualizations}

Polystore applications,
visualizations and clients may need to interact with disparate
database and storage engines. Through the BigDAWG API and middleware,
these applications can use a single interface to any number of
underlying systems. In order to minimize the impact to existing
applications, the ``island'' interface allows users to develop
their applications using the language(s) or data model(s) that most
efficiently (or easily) represents the queries or analytics they are
developing (or have already developed). For example, an application developed using SQL can
leverage the relational island or
a scientific application can leverage the array island. In both cases,
the applications may talk to the same underlying data engines.

In our current implementation, BigDAWG supports a variety of visualization platforms such as
Vega~\cite{satyanarayan2016reactive} and
D3~\cite{bostock2011d3}. Most recently, applying BigDAWG 
to the MIMIC II dataset allowed for the development of a number of
polystore applications:

\begin{enumerate}[leftmargin=*]
\item \textbf{Browsing}: This screen provides an interface to the full
  MIMIC II dataset which is stored in different storage engines. This
  screen utilizes the open source tool ScalaR~\cite{battle2013dynamic}.
\item \textbf{``Something interesting''}: This application uses
  SeeDB~\cite{vartak2014seedb} to highlight interesting trends and
  anomalies in the MIMIC II dataset. 
\item \textbf{``Text Analytics''}: This application performs topic modeling
  of the unstructured doctor and nurse notes directly in a key-value store
  database using Graphulo~\cite{hutchison2015graphulo} and correlates
  them with structured patient information stored in PostGRES.
\item \textbf{``Heavy Analytics''}: This application looks for
  hemodynamically similar patients in a dataset by comparing the
  signatures of historical ECG waveforms using Myria. We discuss this
  particular application in detail in Section~\ref{polystorequery}
\item \textbf{``Streaming Analytics''}: This application performs
  analytics on streaming time-series waveforms and can be used for
  extract-transform-load (ETL) via the data migrator
  into another database such as SciDB.
\end{enumerate}

\section{BigDAWG Performance }
\label{apps}

The current reference implementation of the BigDAWG system satisfies two key performance goals: 1) The
polystore architecture of Figure~\ref{fig:bigdawgarch} can provide low overhead
access to data in disparate engines and 2) Polystore queries can 
outperform ``one size fits all'' solutions. In this section, 
we discuss performance results
with respect to these two goals.

\subsection{BigDAWG overhead}

\begin{figure}[t!]
\centerline{
\includegraphics[width=3.7in]{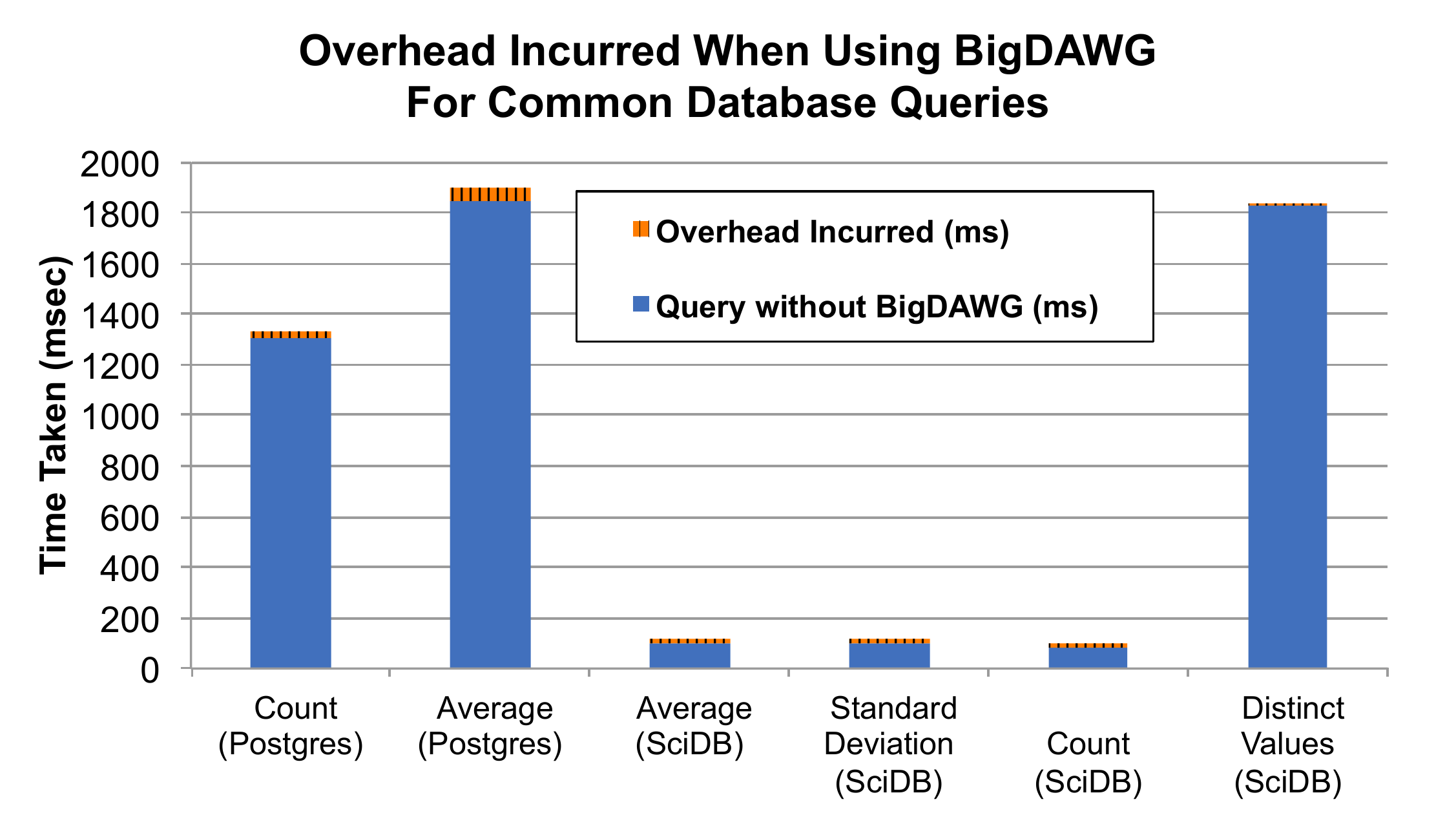}
}
\caption{Overhead of using BigDAWG to execute queries to PostGRES or
  SciDB. For most queries, the overhead is less than 1\% of the
  overall execution time.}
\label{fig:overhead}
\end{figure}

Providing low overhead access to data is an important element of the
BigDAWG system. Low overhead ensures that the BigDAWG middleware and
``island'' architecture do not penalize clients or applications for
using BigDAWG. Supporting low overhead queries is especially important for
applications such as interactive analytics and
visualizations~\cite{ieeeviz2015}.  In Figure~\ref{fig:overhead}, we show the overhead of
executing queries to a single data engine via BigDAWG compared with
the time taken for directly querying the database engine through its
native interface. As we can observe, for most queries, the overhead
incurred by using BigDAWG is a small percentage of the overall query
time. There is a minimum overhead incurred which may be a larger
percentage for queries of shorter duration.

\input{heavy-analytics}

\section{Conclusion and Further Work}
\label{conc}

Future analytics will require access to disparate database management
systems. Previously developed federated and parallel database engines
provided a first step towards the solution but were largely limited to
working with single data or programming models. The concept of
polystore systems extends these concepts to support multiple query
languages and disparate DBMSs. We described our architecture for such
a polystore system, BigDAWG. A reference version of the BigDAWG architecture has been built and
applied to the diverse medical dataset - MIMIC II. Initial performance results validate
that a polystore approach to data management can be applied without
excessive overhead. Further, initial results on a polystore
medical application reinforce the notion that we can achieve greater
performance when using multiple storage engines that are optimized for
particular operations and data models.

There are many areas of potential improvement of the BigDAWG
system. For example, we are interested in developing more complex
query planning and execution capablities, increasing the number of supported
islands and engines, and applying BigDAWG to a greater variety of datasets.

\section*{Acknowledgement}
This work was supported in part by the Intel Science and Technology
Center (ISTC) for Big Data. The authors wish to thank our ISTC collaborators Kristin Tufte, Jeff Parkhurst, Stavros Papadopoulos,
Nesime Tatbul, Magdalena Balazinska, Bill Howe, Jeffrey Heer, David Maier,
Tim Kraska, Ugur Cetintemel, and Stan Zdonik. The authors also wish to
thank the MIT Lincoln Laboratory Supercomputing Center for their help in
maintaining the MIT testbed.

\begin{scriptsize}
\setlength{\parsep}{-1pt}\setlength{\itemsep}{0cm}\setlength{\topsep}{0cm}
\bibliographystyle{IEEEtran}
\bibliography{references.bib}
\end{scriptsize}

\end{document}

%% file: heavy-analytics.tex
%!TEX root = paper.tex

\subsection{Polystore Analytic: Classifying Hemodynamic Deterioration}
\label{polystorequery}

\begin{figure*}[t!]
\centerline{
\includegraphics[width=5.7in]{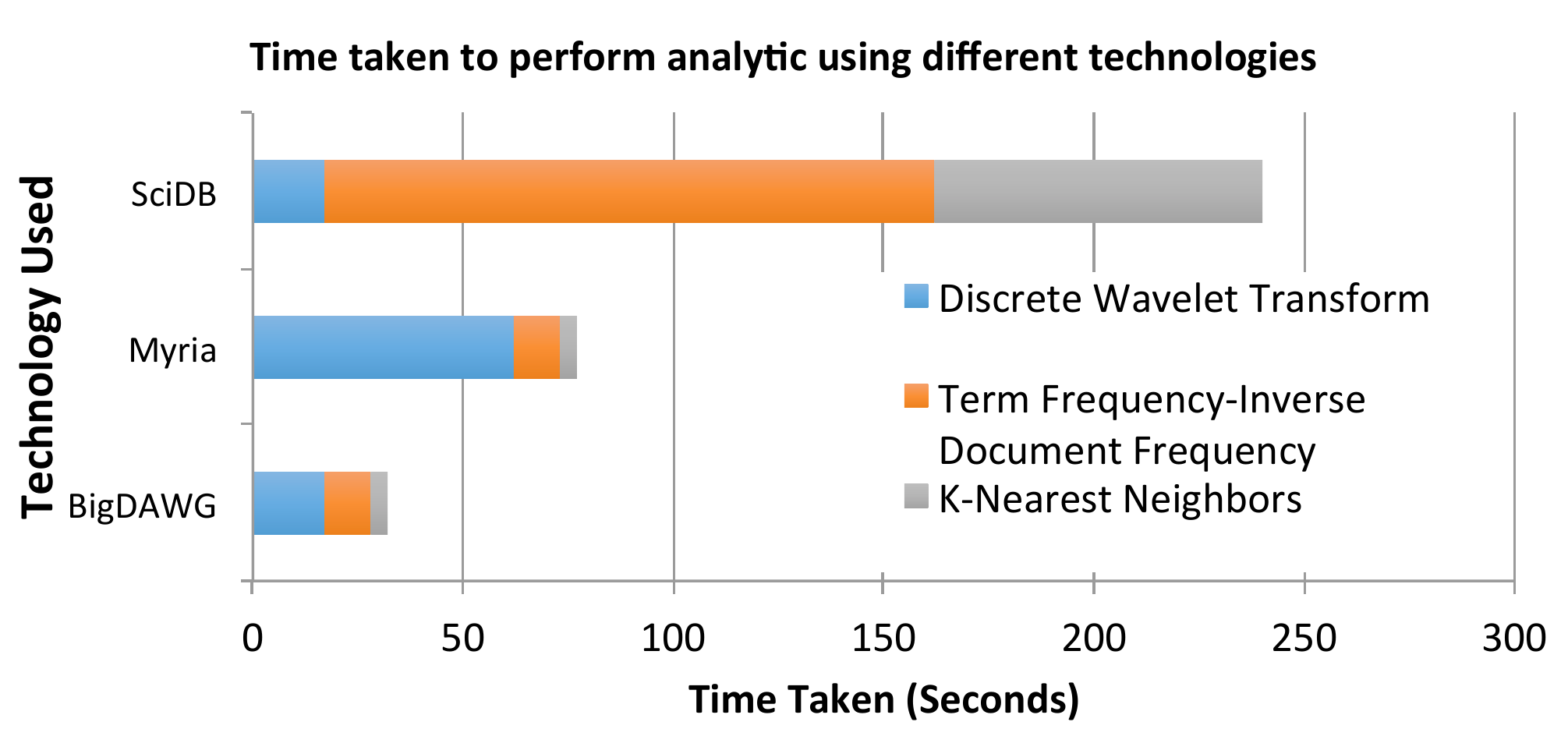}
}
\caption{Polystore analytic applied to medical dataset for 256 minute
  \textsc{ecg} vectors drawn from 600 patients. The polystore
  (Myria+SciDB) execution strategy outperforms a ``one size''
  approach of using just Myria or SciDB. }
\label{fig:polystoreanalytic}
\end{figure*}

To demonstrate the performance advantages offered by a polystore in executing a
complex analytical query, we replicated a process described by Saeed \&
Mark~\cite{saeed2006novel}. This process begins by identifying temporally-similar patterns
in the physiologic measurements of patient data found in the MIMIC II
dataset.  These patterns are used as input to a classifier which identifies
subsequent patients as being likely (or not) to experience hemodynamic
deterioration.

Using the process described by Saeed \& Mark, we first compute the Haar-basis
transform~\cite{haar1910theorie} over the \textsc{ecg} waveforms of training patients.
For each patient, we then binned the coefficients over each temporal scale and
concatenated the resulting histograms into a single patient vector.  We then
normalized each patient vector by applying a term frequency-inverse document
frequency (TF-IDF) computation. Finally, we classified a test patient by
performing a $k$-nearest neighbor computation using these frequency-adjusted
vectors.

We first executed this workflow on our polystore prototype under each of the
Myria and SciDB degenerate islands.  We then performed a multi-island
execution designed to capture performance advantages that exist between each
of these systems.  This execution first computes the Haar-basis transform on
SciDB and casts the intermediate coefficients to Myria, where the TF-IDF and $k
$-NN computation is performed.

We trained a classifier under each configuration using
256-minute \textsc{ecg} vectors drawn from 600 patients present in the MIMIC
II dataset and classified a single test patient.  Each execution was performed
on a cluster comprised of eight \texttt{m4.large} Amazon EC2 instances (\textit{https://aws.amazon.com/}).

As illustrated in Figure~\ref{fig:polystoreanalytic}, we found that
performance under the hybrid configuration (32 seconds) exceed performance
under both the Myria and SciDB islands in isolation (77 and 240 seconds,
respectively).

Our results highlight that substantial performance differences exist between
systems when executing this complex analytical query.  For example,
performance of the TF-IDF and $k$-NN computations are substantially faster in
Myria than SciDB while the wavelet transform time
under SciDB greatly exceeds that of Myria.

For this analytical query, the ability to capture these performance
differences under a polystore yields a substantial performance benefit.  More
generally, our results support the notion that overall query performance may be improved
by identifying and leveraging relative strengths of disparate database systems
within a polystore, and that this improvement far exceeds the cost of
inter-system data casts.